# A Serverless Tool for Platform Agnostic Computational Experiment Management

Gregory Kiar[a,b], Shawn T. Brown[a], Tristan Glatard[*,c], Alan C. Evans[*,a,b,d]

a: Montreal Neurological Institute, McGill University, Montreal, Canada
b: Department of Biomedical Engineering, McGill University, Montreal, Canada
c: Department of Computer Science, Concordia University, Montreal, Canada
d: Department of Neurology and Neurosurgery, McGill University, Montreal, Canada
*: Authors contributed equally
Corresponding author email: <greg.kiar@mcgill.ca>

## Abstract

Neuroscience has been carried into the domain of big data and high performance computing (HPC) on the backs of initiatives in data collection and an increasingly compute-intensive tools. While managing HPC experiments requires considerable technical acumen, platforms and standards have been developed to ease this burden on scientists. While web-portals make resources widely accessible, data organizations such as the Brain Imaging Data Structure and tool description languages such as Boutiques provide researchers with a foothold to tackle these problems using their own datasets, pipelines, and environments. While these standards lower the barrier to adoption of HPC and cloud systems for neuroscience applications, they still require the consolidation of disparate domain-specific knowledge. We present Clowdr, a lightweight tool to launch experiments on HPC systems and clouds, record rich execution records, and enable the accessible sharing of experimental summaries and results. Clowdr uniquely sits between web platforms and bare-metal applications for experiment management by preserving the flexibility of do-it-yourself solutions while providing a low barrier for developing, deploying and disseminating neuroscientific analysis.

---

The increasing adoption of distributed computing, including cloud and high-performance computing (HPC), has played a crucial role in the expansive growth of neuroscience. With an emphasis on big-data analytics, collecting large datasets such as the Consortium for Reliability and Reproducibility [1], UK-Biobank [2], and Human Connectome Project [3] is becoming increasingly popular and necessary. While these datasets provide the opportunity for unprecedented insight into human brain function, their size makes non-automated analysis impractical.

At the backbone of science is the necessity that claims are reproducible. The reproducibility of findings has entered the spotlight as a key question of interest, and has been explored extensively in psychology [4], neuroimaging [5], [6], and other domains [7]. Computational experiments must be re-executable as a critical condition for reproducibility, and this bare minimum requirement becomes increasingly challenging with larger datasets and more complex analyses. When re-executable applications fail to reproduce findings, there is a gray area where the source of errors are often unknown and may be linked to misinterpretation of data, computing resources or undocumented execution details, rather than scientific meaning.

As a result, new tools and standards have emerged to aid in producing more reusable datasets and tools, and thereby more reproducible science. The Brain Imaging Data Structure (BIDS) [8] and the associated BIDS apps [9] prescribe a standard for sharing and accessing datasets, and therefore, increasing the accessibility and impact of both datasets and tools. The Boutiques framework [10] allows for software documentation in a machine-interpretable way, allowing the automation of tool execution and evaluation, and software containerization initiatives such as Docker [11] and Singularity [12] allow this to happen consistently across arbitrary computing environments with minimal burden on the user.

Web-platforms such as OpenNeuro [13], LONI Pipeline [14], and CBRAIN [15] simplify the analysis process further by providing an accessible way to construct neuroscience experiments on commonly-used tools and uploaded-datasets. These systems deploy tools on HPC environments and record detailed execution information so that scientists can keep accurate records and debug their workflows. Tools such as LONI's provenance manager [16], Reprozip [17], and ReCAP [18] capture system-level properties such as system resources consumed and files accessed, where tools supporting the Neuroimaging Data Model (NIDM) [19], a neuroimaging-specific provenance model based on W3C-PROV [20], capture information about the domain-specific transformations applied to the data of interest.

The initiatives enumerated above have synergistic relationships, where each solves a small but significant piece of the larger puzzle that is computational and scientific reproducibility and replicability. However, the learning curve associated with adopting each of these technologies is considerable and leveraging them in an impactful way is difficult. We present Clowdr, a lightweight tool that lowers the barrier on researchers to develop, perform, and disseminate reproducible, provenance-rich neuroscience computation.

## Emergent Technologies in Reproducible Neuroscience

Conducting reproducible analyses in neuroscience requires many complementary facets, building on technologies which are commonly adopted as *de facto* standards.

***(i) Data and code interoperability*** Due in part to its simplicity and active public development community, BIDS [8] has become an increasingly prominent data organization format in neuroimaging. This standard makes use of the Nifti file format [21] and human-readable JSON files to capture both imaging and subject-specific data. An important benefit of this data organization is the ability to launch data processing pipelines in the form of BIDS applications [9], which expose a consistent set of instructions compatible with the data organization. Together, these complementary standards are suitable for performing a large variety of neuroimaging experiments. In contexts where tools have heterogeneous interfaces, or data organizations are custom-built for a particular context, the Boutiques [10], [22] framework allows the rich description of a pipeline such that tool execution, validation, and exploration can be automated.

***(ii) Software virtualization*** While virtual machines have long been used for deploying analysis pipelines with complex dependencies in heterogeneous environments, software containers have recently emerged as lighter-weight alternatives suitable for transient data processing applications. Docker [11] provides this functionality across all major host operating systems, but is often not supported by HPC centers due to security vulnerabilities [23], [24]. Singularity [12] addresses the security risks of Docker, but currently only supports Linux operating systems, filling the niche of containerization on shared computing resources. A detailed comparison in the context of medical imaging can be found in [25].

***(iii) Workflow engines*** For researchers composing custom pipelines, Nipype [26] (Python) and PSOM [27] (GNU Octave or MATLAB) enable the construction of dependency graphs between pipeline components, and allow the deployment either to cluster scheduling software or using multiple threads. These tools contain interfaces for many popular tools and simplify their adoption. They also embed provenance capture, fault-tolerance features, and data tracking to avoid recomputations across similar executions.

***(iv) Provenance*** Building on the W3-PROV [20] standard for data provenance metadata put forth by the World Wide Web Consortium, NIDM [19] is a standard which represents processing and data provenance specific to neuroimaging analyses. While this standard is machine-interpretable and interoperable-by-design, supporting this standard currently requires tight integration with analysis pipelines. In LONI pipeline, a provenance model exists which includes detailed records of data use and file lifecycle [16], which is designed to inform data consumers what types of analyses can be and have been performed with the data in question; this tool is tightly coupled with the LONI pipeline ecosystem. The ReCAP [18] project has been developed to evaluate the resource consumption of arbitrary pipelines on the cloud and can aid in cloud-instance optimization. While this tool has potential for a large impact in designing both cost effective and scalable analyses, there is considerable overhead as it manages executions through a persistent server and workflow engine. While various other libraries exist to monitor some piece of data or compute provenance, Reprozip [17] is perhaps the most exciting as it uniquely captures records of all files accessed and created throughout an execution, which allows for the creation of rich file dependency graphs. The limitation of this technique is that it requires data of interest to be written to disk, as opposed to managed in memory.

***(v) Web Platforms*** Increasing the portability and accessibility of launching large scale analyses, web platforms such as CBRAIN [15], LONI pipeline [14], and OpenNeuro [13] provide science-as-a-service where users can upload and process their data on distant computing resources. Additionally, these platforms provide an accessible and immediate way to share the results produced from experiments with the public. These tools provide incredible value to the community and allow the deployment of production-level pipelines from the web, but they are not suitable for prototyping analyses or developing pipelines, and it is cumbersome to run these services on a lab's own resources. In addition, monolithic Web interfaces are only suitable for a certain type of use-cases and high-level users, while developers or computer-savvy users prefer to rely on modular command-line tools and libraries.

## The Clowdr Microtool

While the technologies enumerated above are essential pieces towards reproducible neuroscience, they are largely isolated from one another and place a large burden on researchers who wish to adopt all of these best practices. Clowdr leverages these tools to increase the deployability, provenance capture, and shareability of experiments. In summary, Clowdr:

**(i)** is tightly based on Boutiques and is BIDS-aware, supporting both arbitrary pipelines and providing an accessible entrypoint for neuroimaging;

**(ii)** executes both bare-metal workflows and Docker or Singularity virtualized pipelines through Boutiques on local, HPC, and cloud resources;

**(iii)** supports the parallelized batch deployment of pipelines constructed with workflow-engines, while being agnostic to programming language and construct;

**(iv)** captures system-level provenance information (i.e. CPU and RAM usage), supports Reprozip, and internal provenance captured by arbitrary pipelines such as NIDM; and

**(v)** supports the deployment of both development- and production-level tools without an active server, and provides a web-report for exploring and sharing experiments.

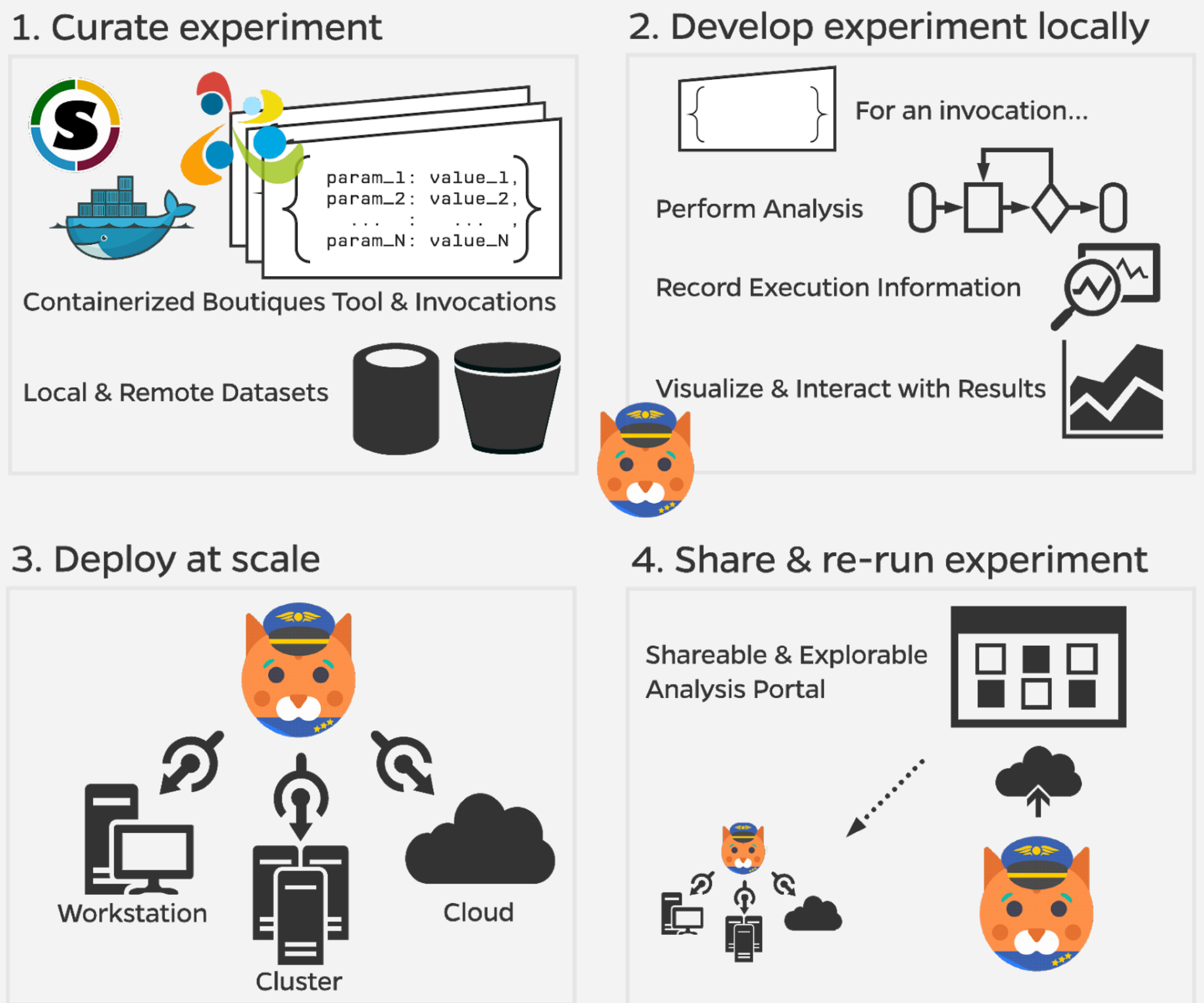

**Figure 1: Clowdr Workflow.** (1) Prior to launching an analysis with Clowdr, users must curate the dataset(s) of interest and analysis tools, like in any analysis. For the sake of portability, Clowdr supports containerized applications described with Boutiques, and both local and Amazon S3-hosted datasets. (2) Clowdr then analyze a subset of the dataset through local-deployments while collecting records of each execution for review, enabling the scientist to document the tuning of their analysis. (3) Clowdr can then be used to deploy experiments on either a local workstation, HPC environment, or an available cloud resource. (4) The metadata collected during these executions can be explored and shared. Together, Clowdr encapsulates an entire experiment reproducibly and with low overhead to the scientist.

A typical workflow using Clowdr is summarized in Figure 1. A Clowdr experiment follows the same workflow as traditional experiments, beginning with tool and data curation through prototyping, deployment, and exploration. Clowdr enhances this experimental process by providing an easy-to-use and lightweight method for launching analyses in distributed, consistent, and richly recorded experimental environments. Clowdr provides graphical reports which can be explored and published alongside the experimental derivatives.

Figure 2 shows the execution lifecycle within Clowdr. Starting from user-provided Boutiques descriptor and invocation(s), and access to any required datasets, Clowdr begins by identifying a list of tasks to launch. For a new experiment, tasks are identified in one of three main ways: 1) a one:one mapping from a list of invocations, 2) a one:many mapping from a single invocation in which parameter(s) have been specified for sweeping during execution, or a BIDS-specific 3) one:many mapping from a single invocation for a BIDS app, which will iterate upon the participant- and session-label fields, and described in the BIDS app specification [9]. Experiments can be re-run and determine the task-list based on whether a full, failure-only, or incomplete-only re-execution is desired. Once the task-list is determined, Clowdr creates an independent invocation which explicitly defines the arguments used in each task.

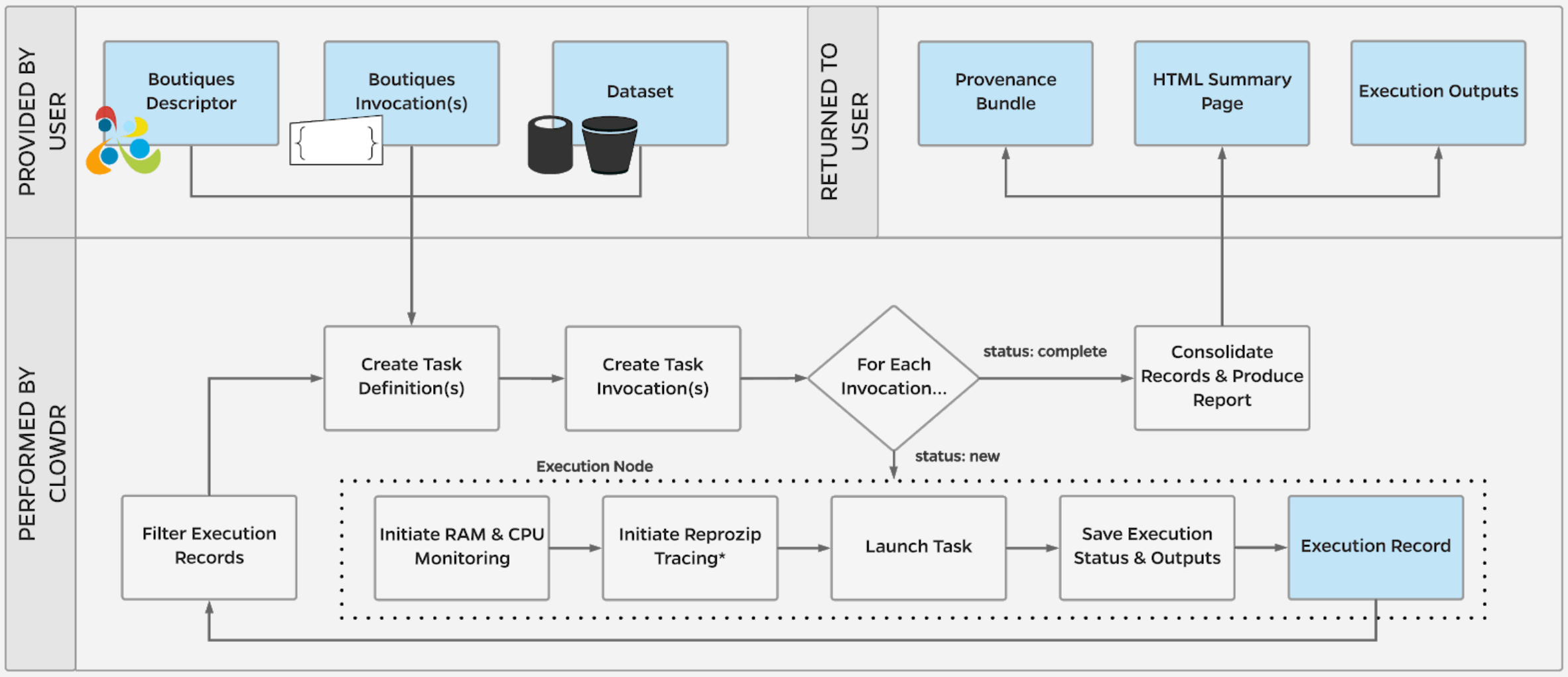

**Figure 2: Clowdr Data Flow.** Beginning with user-supplied tool and data descriptions, Clowdr identifies unique tasks to launch and wraps each with usage and log monitoring tools, to ultimately provide a rich record of execution to the user alongside the expected output products of the experiment. Clowdr ultimately produces an HTML summary for users to explore, update, filter, and share the record of their experiment. In the above schematic, blue boxes indicate data, where gray indicate processing steps. *External reprozip tracing is supported on limited infrastructures, as running virtualized environments within a trace capture requires elevated privileges which may be a security risk on some systems.

At this stage, Clowdr distributes tasks to the Cloud system or local cluster scheduler being used for deployment. Each task is launched through a Clowdr-wrapper, which initializes CPU and RAM monitoring and triggers Reprozip tracing prior to launching the analysis itself. Reprozip tracing has limited support in conjunction with containerized analyses on HPC systems due to potential security issues. Reprozip is built upon the Linux command “ptrace,” which traces processes to monitor or control them. To eliminate the potential risk of using this tool, it is common for systems to disallow the tracing of administrator-level processes. The requirement of limited administrator privileges by Singularity (during the creation of multiple user namespaces) and Docker (for interacting with the daemon) makes encapsulating these tools within the restricted ptrace scope not possible on many shared systems. For more information on the specific conditions in which these technologies can be made to interoperate please view the Github repository for this manuscript, linked below.

Upon completion of the analysis, Clowdr bundles the system monitored records, standard output and error, exit status, and any other information collected by either the tool itself or the Boutiques runtime engine, and concludes its execution. Once the experiment has begun, Clowdr provides the user with the Clowdr provenance directory which will be updated automatically as executions progress.

The researcher can monitor the provenance directory using the Clowdr share portal (Figure 3), which provides a web interface summarizing the task executions. Once the analysis concludes, the figures on this web page and the associated metadata can be saved and serve as a record of the experiment either for evaluation or dissemination alongside published results.

The Clowdr package is open-source on Github [28], and installable through the Python Package Index.

## Performing Experiments with Clowdr

Here we explore an experiment in which we used Clowdr to process the Human Connectome Project (HCP) [3] dataset with a structural and functional connectome estimation pipeline, ndmg [29], [30]. The records of this experiment, and materials and instructions that can be used to reproduce a similar analysis with Clowdr using the publicly available DS114 BIDS dataset [13] an the example BIDS application [9] can be found on Github at: https://github.com/clowdr/clowdr-paper. Specific packages and their versions for both experiments can be found at the end of this manuscript.

As summarized above, performing an analysis with Clowdr requires the creation of a Boutiques descriptor summarizing the pipeline of interest, an invocation containing the parameters to supply to this pipeline on execution, and curation of the data to be processed.

Clowdr experiments can be launched locally, on cluster, or submitted to cloud resources. In each case, invocations and task definitions are created locally, and then the jobs are run serially, submitted to a cluster queue, or pushed to cloud storage and called remotely, respectively. Upon completion of each tasks, summary files created by Clowdr can be either inspected manually or consolidated and visualized in the web with the Clowdr share tool (Figure 3).

| | ▲Task | analysis_level | bids_dir | modality | output_dir | participant_label |
|---|---|---|---|---|---|---|
| | 0 | participant | /data/hcp1200_mir | func | /data/hcp1200_mir | 100206 |
| | 1 | participant | /data/hcp1200_mir | func | /data/hcp1200_mir | 100307 |
| | 2 | participant | /data/hcp1200_mir | func | /data/hcp1200_mir | 100408 |
| | 3 | participant | /data/hcp1200_mir | func | /data/hcp1200_mir | 100610 |
| | 4 | participant | /data/hcp1200_mir | func | /data/hcp1200_mir | 101006 |
| | 5 | participant | /data/hcp1200_mir | func | /data/hcp1200_mir | 101107 |

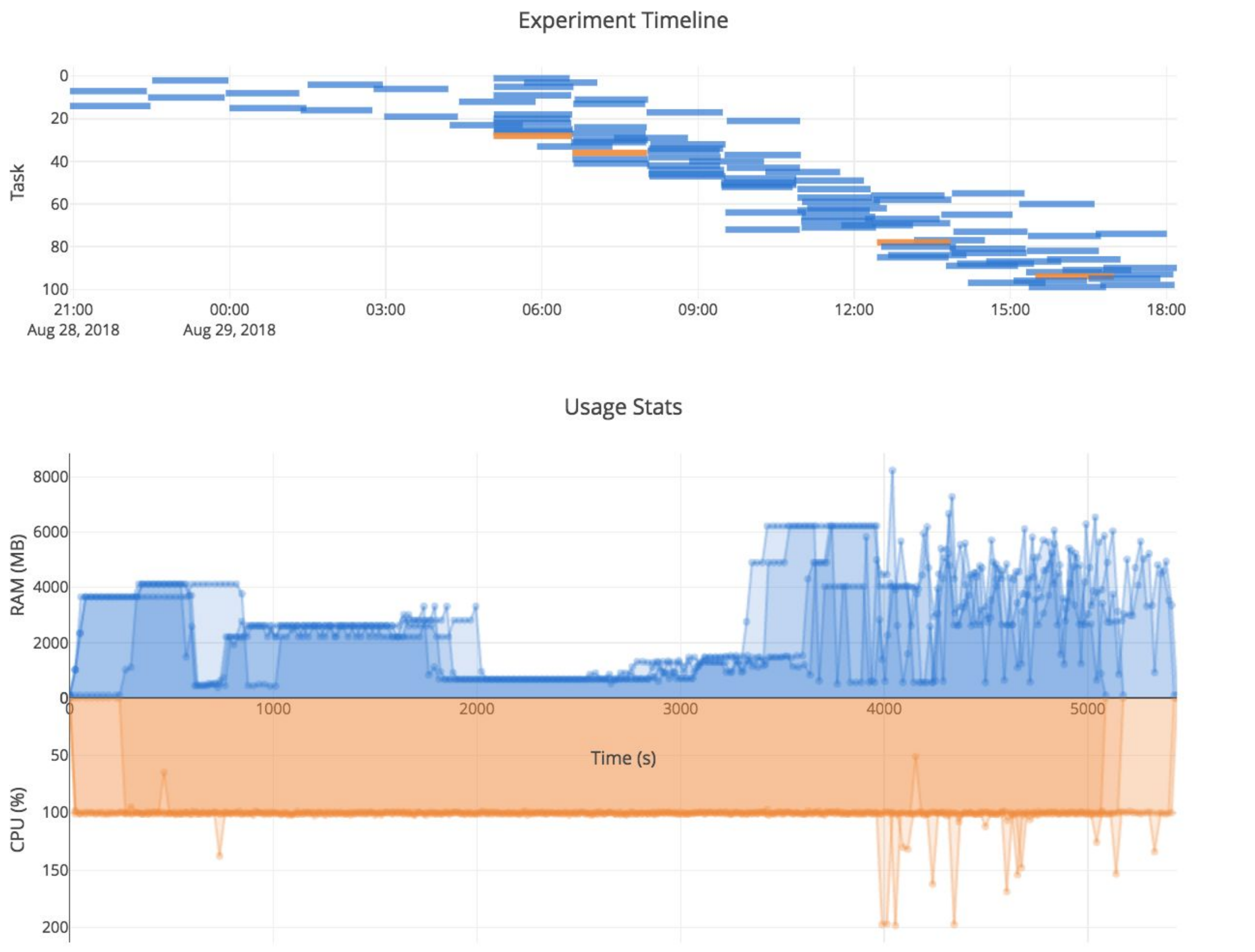

**Figure 3: Clowdr Experiment Viewer.** Once a Clowdr experiment has been created a lightweight service can be launched to monitor experiment progress and outcomes. The page is produced using Plotly Dash to produce highly interactive plots and tables, enabling rich filtering and exploration of executions. The table can be toggled to present summary statistics about experiment execution or invocation parameters identifying parameters used for each task in the experiment. The subsequent Gantt plot shows the timeline of executed tasks in the experiment, where those selected for visualization in the usage plot below are highlighted. The final plot in this view shows the memory and processing footprint throughout all selected tasks. Selection and filtering may be done by value in the tables or selection in the task timeline.

The share tool, launchable on any computer with access to the experiment, creates a lightweight web service displaying summary statistics and invocation information from the experiment, including memory usage, task duration, launch order, and log information. The visualizations provided are filterable and sortable, enabling users to interrogate and identify outliers in their experiment, explore potential sources of failure, and effectively profile the analysis pipeline in use. The modified figures can be downloaded from this interface, serving as accessible records of execution.

In the example above, the HCP dataset has been used to create estimates of structural and functional connectivity. The plot has been filtered to show the usage for the five most RAM-intensive tasks when performing functional connectome estimation. Using these executions as a reference, we are able to identify resource requirements and an experimental trajectory for the ndmg pipeline, and identify outliers in our execution by either filtering explicitly on processing statistics, or comparing across individual trajectories and output or error logs. Clowdr provides a layer of quality control on executions, in addition to that which is regularly performed by researchers on their datasets. This enables automated outlier extraction from an analysis, regardless of the application.

While the share tool currently requires maintaining an active server, the plots can be exported statically and it is in the development roadmap to enable exporting the entire web page as static files, as discussed here: https://github.com/plotly/dash/issues/266. Since the record created by Clowdr is stored in the machine-readable and JSON format, researchers can easily extract their records and integrate it into other interfaces that suit their application.

## Discussion

Clowdr addresses several barriers to performing reproducible neuroscience. Clowdr experiments consist of enclosed computational environments, versioned-controlled Boutiques-described tools with explicit usage parameters, rich execution history, and can be re-executed or distributed with minimal effort. Clowdr provides an accessible interface for initially running analyses locally, and translating them seamlessly to HPC environments. The rich record keeping provided with Clowdr is system-agnostic resulting in uniformly interpretable summaries of execution. As a Python library, Clowdr can be used as a module in a larger platform, or directly as a command-line tool.

Clowdr uniquely packages an executable tool summary, parameters, and results together, in a language- and tool-agnostic way, and therefore, greatly increasing the transparency and shareability of experiments. Importantly, this adds clarity to experimental failures and documents the hyper-parameter tuning process of experiments, which has been historically largely undocumented in literature [31].

Where workflow systems such as Nipye and PSOM provide rich provenance records and enable complex construction of processing pipelines, Clowdr maintains the ability to wrap these as “black box” tools. This permits users to use these tools and benefit from their architecture without requiring knowledge of the underlying language constructs being used in these systems. Platforms like CBRAIN provide a similar type of abstraction, where tools are treated as single objects, but these systems come with the added overhead of maintaining complex database architectures, and primarily accessing resources through a web-based interface. An additional limitation of large platforms is that they are often designed for consumers of widely-adopted tools consumers rather than tool developers. Clowdr fills the void between these two extremes of pipeline deployment systems by providing a programmatic tool-independent method for managing job submission and collecting provenance across multiple architectures and enabling the rapid prototyping of analyses.

Several immediate applications of the provenance information captured by Clowdr include the benchmarking of tools and resource optimization during the selection of cloud resources, as was done in [18]. While the value of comparing provenance records has not been demonstrated here, other studies such as [32] have demonstrated the efficacy of leveraging provenance information to identify sources of variability or instability within pipelines.

Future work includes adopting a W3C-PROV compatible format for Clowdr provenance records, increasing the machine-readability and interoperability of these records with other standards such as NIDM. Integrating the reports produced by Clowdr with a system such as Datalad would allow for record versioning and more strictly enforce the complete reporting of experiments. Clowdr will also continually be extended to support more HPC schedulers, clouds, and provenance capture models.

## Acknowledgements

Funding for this work was provided by The Canada First Research Excellence Fund, Healthy Brains for Health Lives, and The Natural Sciences and Engineering Research Council of Canada, and the Canadian Institutes of Health Research. The authors would also like to thank Pierre Rioux for his insight and many helpful discussions. The authors declare no conflicts of interest.

## Tools and Versions

The following is a list of tools and data used in this manuscript, and their respective versions. The architecture and analysis presented for the Clowdr package corresponds to version 0.1.0. The key Python packages and specific versions tested are: boutiques (version 0.5.12), boto3 (1.7.81), botocore (1.10.81), slurmpy (0.0.7), psutil (5.4.7), pandas (0.23.4), plotly (3.1.1), and plotly-dash (0.24.1), including dash-core-components (0.27.1), dash-html-components (0.11.0), dash-renderer (0.13.0), dash-table-experiments (0.6.0), and flask (0.12.2). Executions were tested locally using Docker (17.12.0-ce), and on Compute Canada’s Cedar high performance cluster using Singularity (2.5.1-dist). The Docker container used for ndmg can be found on Dockerhub as neurodata/m3r-release (0.0.5), which contains ndmg (0.1.0-f). The Singularity container used was pulled and dynamically created from this Dockerhub endpoint. The dataset use was a subset of the HCP 1200 collection [3].